\documentclass[12pt]{article}
\usepackage{epsfig}
\usepackage{amsmath}
\usepackage{hhline}
\usepackage{amssymb}
\usepackage{times}
\usepackage{cite}

\newlength{\dinwidth}
\newlength{\dinmargin}
\begin{document}  
\newcommand{\pom}{{I\!\!P}}
\newcommand{\reg}{{I\!\!R}}
\newcommand{\slowpi}{\pi_{\mathit{slow}}}
\newcommand{\fiidiii}{F_2^{D(3)}}
\newcommand{\fiidiiiarg}{\fiidiii\,(\beta,\,Q^2,\,x)}
\newcommand{\n}{1.19\pm 0.06 (stat.) \pm0.07 (syst.)}
\newcommand{\nz}{1.30\pm 0.08 (stat.)^{+0.08}_{-0.14} (syst.)}
\newcommand{\fiidiiiful}{F_2^{D(4)}\,(\beta,\,Q^2,\,x,\,t)}
\newcommand{\fiipom}{\tilde F_2^D}
\newcommand{\ALPHA}{1.10\pm0.03 (stat.) \pm0.04 (syst.)}
\newcommand{\ALPHAZ}{1.15\pm0.04 (stat.)^{+0.04}_{-0.07} (syst.)}
\newcommand{\fiipomarg}{\fiipom\,(\beta,\,Q^2)}
\newcommand{\pomflux}{f_{\pom / p}}
\newcommand{\nxpom}{1.19\pm 0.06 (stat.) \pm0.07 (syst.)}
\newcommand {\gapprox}
   {\raisebox{-0.7ex}{$\stackrel {\textstyle>}{\sim}$}}
\newcommand {\lapprox}
   {\raisebox{-0.7ex}{$\stackrel {\textstyle<}{\sim}$}}
\def\gsim{\,\lower.25ex\hbox{$\scriptstyle\sim$}\kern-1.30ex%
\raise 0.55ex\hbox{$\scriptstyle >$}\,}
\def\lsim{\,\lower.25ex\hbox{$\scriptstyle\sim$}\kern-1.30ex%
\raise 0.55ex\hbox{$\scriptstyle <$}\,}
\newcommand{\pomfluxarg}{f_{\pom / p}\,(x_\pom)}
\newcommand{\dsf}{\mbox{$F_2^{D(3)}$}}
\newcommand{\dsfva}{\mbox{$F_2^{D(3)}(\beta,Q^2,x_{I\!\!P})$}}
\newcommand{\dsfvb}{\mbox{$F_2^{D(3)}(\beta,Q^2,x)$}}
\newcommand{\dsfpom}{$F_2^{I\!\!P}$}
\newcommand{\gap}{\stackrel{>}{\sim}}
\newcommand{\lap}{\stackrel{<}{\sim}}
\newcommand{\fem}{$F_2^{em}$}
\newcommand{\tsnmp}{$\tilde{\sigma}_{NC}(e^{\mp})$}
\newcommand{\tsnm}{$\tilde{\sigma}_{NC}(e^-)$}
\newcommand{\tsnp}{$\tilde{\sigma}_{NC}(e^+)$}
\newcommand{\st}{$\star$}
\newcommand{\sst}{$\star \star$}
\newcommand{\ssst}{$\star \star \star$}
\newcommand{\sssst}{$\star \star \star \star$}
\newcommand{\tw}{\theta_W}
\newcommand{\sw}{\sin{\theta_W}}
\newcommand{\cw}{\cos{\theta_W}}
\newcommand{\sww}{\sin^2{\theta_W}}
\newcommand{\cww}{\cos^2{\theta_W}}
\newcommand{\trm}{m_{\perp}}
\newcommand{\trp}{p_{\perp}}
\newcommand{\trmm}{m_{\perp}^2}
\newcommand{\trpp}{p_{\perp}^2}
\newcommand{\alp}{\alpha_s}

\newcommand{\alps}{\alpha_s}
\newcommand{\sqrts}{$\sqrt{s}$}
\newcommand{\LO}{$O(\alpha_s^0)$}
\newcommand{\Oa}{$O(\alpha_s)$}
\newcommand{\Oaa}{$O(\alpha_s^2)$}
\newcommand{\PT}{p_{\perp}}
\newcommand{\JPSI}{J/\psi}
\newcommand{\sh}{\hat{s}}
\newcommand{\uh}{\hat{u}}
\newcommand{\MP}{m_{J/\psi}}
\newcommand{\PO}{I\!\!P}
\newcommand{\xbj}{x}
\newcommand{\xpom}{x_{\PO}}
\newcommand{\ttbs}{\char'134}
\newcommand{\xpomlo}{3\times10^{-4}}  
\newcommand{\xpomup}{0.05}  
\newcommand{\dgr}{^\circ}
\newcommand{\pbarnt}{\,\mbox{{\rm pb$^{-1}$}}}
\newcommand{\gev}{\,\mbox{GeV}}
\newcommand{\WBoson}{\mbox{$W$}}
\newcommand{\fbarn}{\,\mbox{{\rm fb}}}
\newcommand{\fbarnt}{\,\mbox{{\rm fb$^{-1}$}}}
%
%
\newcommand{\qsq}{\ensuremath{Q^2} }
\newcommand{\gevsq}{\ensuremath{\mathrm{GeV}^2} }
\newcommand{\et}{\ensuremath{E_t^*} }
\newcommand{\rap}{\ensuremath{\eta^*} }
\newcommand{\gp}{\ensuremath{\gamma^*}p }
\newcommand{\dsiget}{\ensuremath{{\rm d}\sigma_{ep}/{\rm d}E_t^*} }
\newcommand{\dsigrap}{\ensuremath{{\rm d}\sigma_{ep}/{\rm d}\eta^*} }
\def\Journal#1#2#3#4{{#1} {\bf #2} (#3) #4}
\def\NCA{Nuovo Cimento}
\def\NIM{Nucl. Instrum. Methods}
\def\NIMA{{Nucl. Instrum. Methods} {\bf A}}
\def\NPB{{Nucl. Phys.}   {\bf B}}
\def\NPPS{Nucl. Phys. Proc. Suppl.} 
\def\NPPSC{{Nucl. Phys. Proc. Suppl.} {\bf C}}
\def\PLB{{Phys. Lett.}   {\bf B}}
\def\PRL{Phys. Rev. Lett.}
\def\PRD{{Phys. Rev.}    {\bf D}}
\def\ZPC{{Z. Phys.}      {\bf C}}
\def\EJC{{Eur. Phys. J.} {\bf C}}
\def\CPC{Comp. Phys. Commun.}
\def\NP{{Nucl. Phys.}}
\def\JPG{{J. Phys.} {\bf G}} 
\def\EPC{{Eur. Phys. J.} {\bf C}}
\def\JHEP{{JHEP} { }}

\begin{titlepage}
\begin{figure}[!t]
DESY 04--051 \hfill ISSN 0418--9833 \\
April 2004
\end{figure}

\bigskip
\vspace*{2cm}

\begin{center}
\begin{Large}
 
{\bf Forward \boldmath $\pi^\circ$ Production and Associated Transverse Energy 
Flow in Deep-Inelastic Scattering  at HERA}

\vspace{2cm}

H1 Collaboration

\end{Large}
\end{center}

\vspace{2cm}

\begin{abstract}
Deep-inelastic positron-proton
interactions at low values of Bjorken-$x$ down to $x \approx 4\cdot
10^{-5}$ which give rise to high transverse momentum $\pi^\circ$-mesons
are studied with the H1 experiment at HERA.
The inclusive cross section for $\pi^\circ$-mesons produced at small 
angles
with respect to the proton remnant (the forward region) is presented
as a function of the transverse momentum and energy of the $\pi^\circ$  
and of the four-momentum transfer $Q^2$ and Bjorken-$x$.  
Measurements are 
also presented of the transverse energy flow in events containing a 
forward $\pi^\circ$-meson.
Hadronic final state calculations based on QCD models implementing 
different parton evolution schemes are confronted with the data.

\noindent

\end{abstract}

\vspace{1.5cm}

\begin{center}
To be submitted to Eur. Phys. J. C
\end{center}

\end{titlepage}

\begin{flushleft}


A.~Aktas$^{10}$,               
V.~Andreev$^{24}$,             
T.~Anthonis$^{4}$,             
A.~Asmone$^{31}$,              
A.~Astvatsatourov$^{35}$,      
A.~Babaev$^{23}$,              
S.~Backovic$^{35}$,            
J.~B\"ahr$^{35}$,              
P.~Baranov$^{24}$,             
E.~Barrelet$^{28}$
% ...

\bigskip{\it
% ...
$ ^{4}$ Inter-University Institute for High Energies ULB-VUB, Brussels;
          Universiteit Antwerpen, Antwerpen; Belgium$^{ c}$ \\
% ...
$ ^{10}$ DESY, Hamburg, Germany \\
% ...
$ ^{23}$ Institute for Theoretical and Experimental Physics,
          Moscow, Russia$^{ l}$ \\
 $ ^{24}$ Lebedev Physical Institute, Moscow, Russia$^{ e}$ \\
% ...
$ ^{28}$ LPNHE, Universit\'{e}s Paris VI and VII, IN2P3-CNRS,
          Paris, France \\
% ...
$ ^{31}$ Dipartimento di Fisica Universit\`a di Roma Tre
          and INFN Roma~3, Roma, Italy \\
% ...
$ ^{35}$ DESY, Zeuthen, Germany \\
% ...
$ ^c$ Supported by FNRS-FWO-Vlaanderen, IISN-IIKW and IWT and by 
       Inter-University Attraction Poles Programme, Belgian Science Policy \\
% ...
$ ^e$ Supported by the Deutsche Forschungsgemeinschaft \\
% ...
$ ^l$ Partially Supported by Russian Foundation
      for Basic Research, grant    no. 00-15-96584 \\
}

\vspace{2cm}

\begin{center}
  Dedicated to the memory of Krzysztof Rybicki.
\end{center}

\end{flushleft}

\newpage

\section{Introduction}

Measurements of the hadronic final state in deep-inelastic
positron-proton scattering (DIS) at HERA 
have allowed precision 
tests of the theory of the strong force, Quantum Chromodynamics
(QCD). In particular, properties of the hadronic final state in the 
region near to the 
proton remnant system (hereafter referred to as the `forward region') have 
been shown to be sensitive to  
the QCD radiation pattern formed from the cascade initiated by a 
parton from the proton before it undergoes a hard 
scatter\cite{fjet1,fjet2,zeusfj,et1}. This paper presents studies made 
by the H1 experiment of 
DIS interactions at values of Bjorken-$x$ down to $x\approx 4\cdot 
10^{-5}$ containing at least one high transverse momentum, forward going 
$\pi^\circ$-meson.

\begin{figure}[htb]
\begin{center}
\epsfig{file=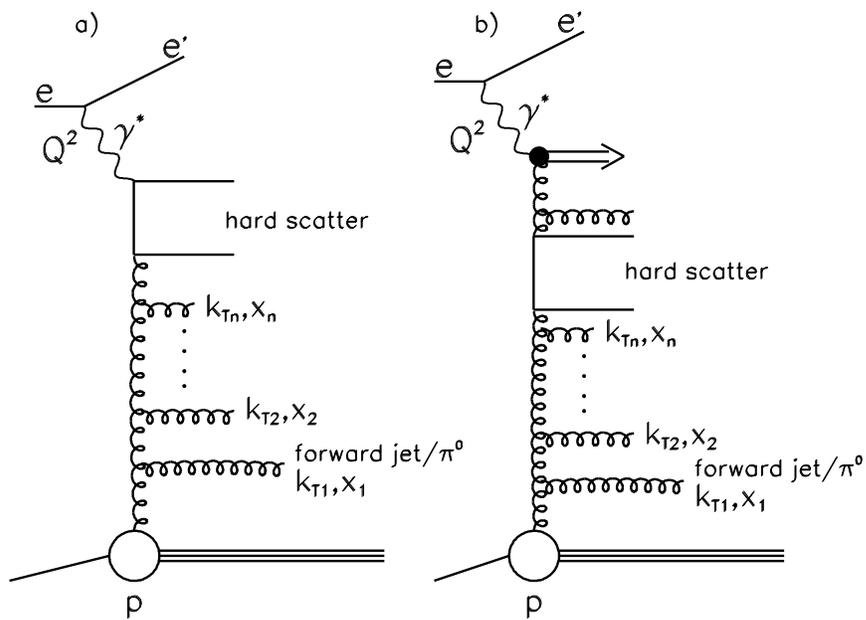,width=14.2cm,height=14.2cm}
\caption{Generic diagrams for DIS processes at small $x$. 
The gluon longitudinal
momentum fractions and transverse momenta are labelled $x_i$ and $k_{Ti}$,
respectively.
(a) A gluon ladder evolves between the quark box, attached to the virtual 
photon, and the proton. (b) The partonic structure of the photon is ``resolved''.}
\label{fig0}
\end{center}
\end{figure}

A generic diagram for parton evolution in a DIS process at low $x$ in 
which a gluon from the proton undergoes a QCD cascade is shown in 
Fig.~\ref{fig0} (a). The gluon 
eventually interacts with the virtual photon via a hard photon-gluon 
fusion process which can be calculated within perturbative QCD using 
an exact matrix element.  
Several perturbative QCD-based prescriptions are available to describe the
dynamics of the parton evolution process. The Dokshitzer-Gribov-Lipatov-Altarelli-Parisi
(DGLAP) evolution equations\cite{DGLAP} resum leading $\log(Q^2)$ terms 
and  
ignore $\log(1/x)$ terms. In an axial gauge this corresponds to the 
resummation of diagrams in which the parton cascades follow a strong 
ordering in 
transverse momenta $k_{Tn}^2 \gg k_{Tn-1}^2 \gg \cdots \gg k_{T1}^2$. At 
sufficiently small values of $x$ the Balitsky-Fadin-Kuraev-Lipatov
(BFKL) equation\cite{BFKL} should be applicable since 
the $\log(1/x)$ terms should dominate the evolution. In this scheme 
the cascade is ordered strongly in fractional 
momenta $x_{n} \ll x_{n-1}\ll \cdots \ll x_1$,  
while the transverse momenta perform a `random walk' with 
$k_{Ti}$ being close to $k_{T{i-1}}$, though it can be both larger or 
smaller. The Ciafaloni-Catani-Fiorani-Marchesini (CCFM) equation~\cite{CCFM}  
interpolates between the DGLAP and BFKL approximations with parton emissions 
ordered in angle. 

A parton chain without a requirement of $k_T$ ordering throughout the 
complete cascade is provided
 in a  picture of low $x$ DIS in which the virtual photon is ascribed 
a partonic structure. This process is illustrated in Fig.~\ref{fig0} (b)  
 in which $k_T$-ordered DGLAP cascades are initiated 
both from the proton and photon, leading to the 
hard interaction at the centre of the QCD `ladder'.

Hadronic final state observables are sensitive to the dynamics of QCD 
 processes and are thus expected to be able to discriminate between different 
evolution 
approximations. As illustrated in Fig.~\ref{fig0}, the selection of  
leading particles or jets in the 
forward region can tag individual parton emissions at high 
transverse momentum.
An advantage of using single particles rather than jets is that 
ambiguities due to the choice of jet algorithm are removed. On the other 
hand, uncertainties due to hadronisation are typically larger for single 
particle 
studies. 
As was 
pointed out in\cite{mueller}, the selection 
of particles or jets with values of transverse momentum squared of similar 
magnitude to $Q^2$ suppresses the contribution of $k_T$-ordered cascades 
with respect to $k_T$-unordered processes. In addition, the phase space for 
BFKL effects is enhanced if the fraction of the proton's energy of the 
particle or jet is required to be greater than Bjorken-$x$. 
 
A recent H1 study of
forward going $\pi^\circ$-mesons\cite{fpi2} found that a model which 
implemented 
a DGLAP parton cascade from the proton
significantly underestimated the cross section 
at low values of Bjorken-$x$. 
Leading order BFKL calculations with kinematic constraints which  
mimic higher orders and a model implementing virtual photon 
structure gave a better description of the data.  

The pseudorapidity dependence of the mean transverse energy  
produced in an event (the so-called transverse energy flow) 
provides a complementary means of studying the hadronic final state.
Compared with studies of jets and high transverse momentum particles, 
measurements of transverse  
energy flow typically cover a wider range of pseudorapidity and are 
sensitive to parton emissions of lower transverse momentum. 
Previous measurements of 
transverse energy flow at HERA were indeed found to be sensitive to the 
modelling of both the perturbative QCD evolution and the soft 
hadronisation process\cite{etzeus,et1,et3}. Measurements of transverse 
energy flow in events containing 
particles with high 
transverse momentum also reveal the range over which 
transverse momentum is compensated following the emission of QCD 
radiation~\cite{emc}. 

This paper presents a study of low $x$ DIS interactions 
in which high transverse momentum
$\pi^\circ$-mesons are produced in the forward region. The results 
are 
based on a data sample which is more than three times 
larger than that used for earlier studies\cite{fpi2}.
Consequently, the inclusive 
$\pi^\circ$ 
cross section is measured with greater precision and more 
differentially as a function of $x$, $Q^2$ and the transverse momentum and 
energy of the $\pi^\circ$-meson. Furthermore, for the first time, 
measurements are presented of transverse 
energy flow for $ep$ interactions containing forward going 
$\pi^\circ$-mesons. 
This allows a more complete 
investigation of hadronic final states containing a hard forward 
$\pi^\circ$ than was 
previously possible.

\section{QCD-based Models}
Calculations of the expected production rate of $\pi^\circ$-mesons  and 
the associated transverse energy flow in DIS are available in the form of 
Monte Carlo
event generators. These use first-order QCD matrix 
elements and adopt various approaches 
to modelling the parton cascade. These models are used to  
provide comparisons of theory predictions with the measurements presented 
here. In addition they  are  used, together with a 
simulation of the H1 detector, to correct the measurements for 
the finite acceptance and resolution of the detector.
Unless otherwise stated the proton and virtual photon parton densities used in these models 
are  
{\sc CTEQ6M}\cite{CTEQ6M} and {\sc SAS-1D}\cite{sasg}, respectively.

{\sc Lepto} 6.51\cite{lepto} matches  first-order QCD
matrix
elements to DGLAP-based leading-log parton showers. The factorisation and 
renormalisation 
scales are set to $Q^2$. 
{\sc Lepto} also allows
for non-perturbative rearrangement of the event colour topology 
via so-called soft color interactions in the final state\cite{sci}.

{\sc Rapgap} 2.08/20\cite{rapgap} also matches 
first-order QCD matrix
elements for direct photon processes to DGLAP-based leading-log parton showers. In addition
to the direct photon processes, {\sc Rapgap}
simulates resolved photon interactions in which the virtual photon is
assumed to have partonic structure. For the predictions
presented here, the renormalisation  
and factorisation scales are set to
$4p_T^2 + Q^2$, where $p_T$ is  the transverse momentum of 
the partons emerging from 
the hard scattering process. The hadronic final state predictions of {\sc Rapgap}, when only direct 
photon interactions are considered, are very similar to those of {\sc 
Lepto}.

{\sc Ariadne} 4.10\cite{ariadne}  
is an implementation 
of the Colour Dipole Model (CDM) \cite{cdm} of a chain of independently 
radiating dipoles formed by emitted gluons. Since all radiation is assumed 
to come from the dipole formed by the struck quark and the remnant, 
photon-gluon fusion events have to be added and are taken from the QCD 
matrix 
elements. The parameters in {\sc Ariadne} have been optimised 
in order to describe a range of hadronic final state 
measurements\cite{heraws}. 

In order to study the effects of initial and final state QED radiation, 
the above models are 
interfaced with {\sc  Heracles}\cite{heracles} within 
the {\sc Django}\cite{DJANGO} model.

{\sc Cascade} 1.0\cite{cascade} uses off-shell QCD matrix elements, 
supplemented with parton emissions based on the CCFM equation within a 
backward 
evolution approach. An unintegrated gluon density, obtained using CCFM
evolution and fitted
to describe the inclusive
DIS cross section \cite{cascade1}, is used as an input to this model. In 
the 
present analysis an updated version \cite{newcascade} of 
{\sc Cascade} 
with an 
improved treatment of the soft region and a new parameterisation of the
unintegrated gluon density is used. These modifications provide  
an improved description of forward jet production\cite{smallX}.

To perform the hadronisation step,  all of the above models use  
the {\sc Lund} string fragmentation\cite{string} scheme, as implemented 
in {\sc Jetset}\cite{jetset74} in case of {\sc Lepto}, {\sc Rapgap} and 
{\sc Ariadne}
and in {\sc Pythia}\cite{jetset74} for {\sc Cascade}.

Predictions of the $\pi^\circ$ cross sections are 
also available from an analytical calculation at the 
parton level\cite{outhwait} based on  a modified BFKL 
evolution equation at lowest order. These calculations are then 
convoluted
with a  
$\pi^\circ$ fragmentation function\cite{pi0frag}. The proton parton  
densities are taken from \cite{bfklpar}.
The modified evolution equation imposes 
a ``consistency constraint"\cite{con1,con2} which, 
it is argued, mimics much of the 
contribution from  non-leading $\log(1/x)$ terms to the BFKL equation. 
However, these predictions 
are very sensitive to the choice of scale for the strong coupling constant 
$\alpha_s$ and to the infra-red 
cut-off. 
In the present analysis, the scale for   
$\alpha_s$ is taken to be the squared transverse momentum of the emitted 
partons, $k_T^2$,  
and the infrared cut-off in the modified BFKL equation is set at $0.5$ 
GeV$^2$.\\ 

Recently, calculations of the cross section for the production of high 
transverse momentum hadrons in DIS interactions have been 
made\cite{aurenche}, which describe earlier 
measurements of forward $\pi^\circ$ production\cite{fpi2}. 
These comprise next-to-leading order (NLO) matrix elements, convoluted 
with NLO fragmentation functions\cite{pi0frag}. 
The proton parton densities MRST99 (higher gluon)\cite{MRST99} are used. 
For comparison with the
H1 data, the renormalisation, factorisation and 
fragmentation 
scales are each set to
$(Q^2+p_{T,\pi}^{*2})/2$, where $p_{T,\pi}^*$ is the transverse momentum, 
in the photon-proton centre of mass system\footnote{All quantities 
presented in the hadronic centre-of-mass system are                       
denoted by the superscript *.},  
of the parton which fragments into the forward  $\pi^\circ$. 
In these calculations, a large part of the cross
section is generated by higher order contributions which 
correspond to lowest order BFKL and resolved photon 
processes\cite{aurenche}.

\section{Experimental Apparatus}

A detailed description of the H1 detector can be found elsewhere\cite{h1det}. The following section
briefly describes the components of the detector which are most relevant for this analysis.

A liquid argon (LAr) calorimeter is used to measure 
the hadronic energy flow and the candidate $\pi^\circ$-meson properties. 
The LAr   
calorimeter provides measurements 
over the laboratory polar angle range $4^\circ < \theta < 154^\circ$, 
where $\theta$ 
is defined with respect to the direction of the proton beam,
and offers full azimuthal coverage. It consists of an
electromagnetic section with lead absorbers and a hadronic section with steel absorbers. 
Both sections are highly segmented in the transverse and longitudinal
directions.
The total depth of both sections varies between 4.5 and 8 interaction lengths in the
region $4^\circ < \theta < 128^\circ$, and between 20 and 30 radiation lengths in the region
 $4^\circ < \theta < 154^\circ$ increasing towards the forward direction. 
The fine granularity of the electromagnetic
section in the forward direction is characterized by four-fold longitudinal
segmentation and a typical lateral cell size of 3.5 x 3.5 cm$^2$.  
Test beam measurements of the LAr calorimeter modules
showed an energy resolution of 
$\sigma_E/E  \approx 0.50/\sqrt{E[\rm GeV]} \oplus 0.02$
for charged pions and of $\sigma_E/E 
\approx 0.12/\sqrt{E[\rm GeV]} \oplus 0.01$ for
electrons\cite{h1det}. The hadronic energy measurement is 
made by applying a weighting
technique to the electromagnetic and hadronic components of the energy 
deposition, in order to 
account for the non-compensating 
nature of the calorimeter. The absolute scales of hadronic  
and forward-reconstructed electromagnetic energies are known to 
4\%\cite{echpi} and 3\%\cite{enpi0}, respectively.

The SPACAL is a lead/scintillating fibre calorimeter covering the region $153^\circ <
\theta < 177.5^\circ$ with an electromagnetic and a hadronic section. It is used to 
measure the scattered positron energy and the backward hadronic energy 
flow. The 
energy 
resolutions  
for 
electrons and hadrons are $\sigma_E/E \approx
0.07/\sqrt{E [\rm GeV] } \oplus 0.01 $\cite{SPACALTEST} and 
$\sigma_E/E \approx 0.3/\sqrt{E\ [\rm GeV] }$\cite{shad}, respectively.
The energy scale uncertainties are 1\% for electrons and 7\% 
for hadrons\cite{glazov}.  

The calorimeters are surrounded by a superconducting solenoid which provides a uniform magnetic
field of 1.15 T in a direction parallel to the proton beam in the tracking region. Charged particle
tracks are measured in the central tracker (CT) and forward tracker (FT) systems which cover the
polar angle ranges of 
$25^\circ < \theta < 155^\circ$ and $5^\circ < \theta < 25^\circ$, 
respectively.
Information from the CT is used in this work to trigger events, to locate 
the event vertex and also contributes to the measurement of transverse energy. 

A backward drift chamber   
(BDC) in front of the SPACAL with an angular acceptance of $151^\circ <
\theta < 177.5^\circ$ serves to identify 
electron candidates and to precisely measure their
direction. Using information from the BDC, the SPACAL and the 
reconstructed event vertex position, the
 polar angle of the scattered electron is known to about 
0.5 mrad\cite{glazov}.

The luminosity is measured using the Bethe-Heitler process 
$ep \rightarrow ep\gamma$ with two TlCl/TlBr crystal calorimeters 
installed in the HERA tunnel. 

\section{Data Analysis}
The data used for this analysis were collected in 
1996 and 1997 when positrons  
and protons with energies of $27.6$ GeV and $820$ GeV, respectively, were 
collided. The data-set used in this work corresponds to an 
integrated luminosity of 21.2 pb$^{-1}$. 

\subsection{DIS Event Selection}
DIS events are selected by triggers based on electromagnetic energy 
deposits in the SPACAL calorimeter and the presence of charged particle tracks 
in the CT. For the $\pi^\circ$-enriched event sample, the trigger 
efficiency lies 
between 60\% and 80\%, determined using independently 
triggered data. The inefficiency is mainly due to the 
suppression by the trigger of events 
with less than 3 charged particles measured in the CT. The data are 
corrected for the trigger inefficiency by applying a weight to every 
selected
DIS event.

The event kinematics are calculated from the polar angle
and the energy of the scattered positron.
In order to maintain optimal efficiency and acceptance, scattered positron candidates in 
the SPACAL are required to have an energy 
$E_{e'} > 10 $ GeV and to lie in the region of polar angle of $156^{\circ} 
< 
\theta_{e'} 
< 177^{\circ}$. The data-set is further restricted to the kinematic 
range in inelasticity $y$ and virtuality $Q^2$ of 
$0.1 < y < 0.6$ and $2 < Q^{2} < 70$ GeV$^{2}$, respectively. 
The resulting values of Bjorken-$x$ extend over two orders of magnitude 
in the range  $4\cdot 10^{-5}< x < 6 \cdot10^{-3}$.    

To further reject background from photoproduction interactions in which  
a particle from the hadronic final state is misidentified as a scattered 
positron in the SPACAL, the 
condition $35 < \Sigma_j (E_j - 
p_{z,j}) < 70$ GeV is applied. Here $E_j$ and $p_{z,j}$ are the energy and 
longitudinal 
momentum, respectively, of a particle, and the sum extends 
over all particles in the event 
except those detected in the luminosity system. 

\subsection{Forward \boldmath$\pi^\circ$ Selection}
 The selection of forward $\pi^\circ$-mesons closely follows that 
of earlier work\cite{fjet2,fpi2,twthesis}.
The $\pi^\circ$ candidates are identified via the dominant decay channel 
$\pi^\circ
\rightarrow 2\gamma$ using calorimetric information only. Criteria are placed on the 
kinematic properties of the candidate to ensure high acceptance and 
efficient background rejection. 
The energy $E_\pi$ of the $\pi^\circ$ 
scaled with the energy $E_p$ of the proton beam, $x_\pi=E_\pi/E_p$,  
is required to be greater than 0.01.
As a consequence of this cut, the decay photons are not resolved 
individually but are merged into a 
single
electromagnetic cluster in the detector.   
The candidates are also required to lie in the polar angle region  
$5^{\circ} < \theta_{\pi} <25^{\circ}$. This polar angle 
region,                       
referred to as `forward' in the laboratory frame, corresponds  
to                
the central region in the hadronic centre of mass frame, $-1.25 \ 
\lapprox \ \eta_{\pi}^* \ \lapprox \ 2.0$.
Furthermore, the $\pi^\circ$ transverse 
momentum in the photon-proton centre of mass system, 
$p_{T,\pi}^*$ must be greater than $2.5$ GeV.
The Lorentz boost to this 
frame is calculated using the kinematics of the scattered electron. In 
this frame the proton direction is chosen to define the negative $z^*$ 
axis. 

Electromagnetic and hadronic showers are discriminated by an analysis of the longitudinal 
and transverse shapes of the energy depositions. A $\pi^\circ$ candidate 
is required to have
more than 90\% of its energy deposited in the electromagnetic part of the LAr 
calorimeter. A ``hot'' core of the most energetic group 
of contiguous cells within a 
cluster, which must include the hottest cell, is required to account for 
over $50$\% of the cluster 
energy. The lateral spread of the shower, defined as in \cite{lateral}, is 
required to be 
less than $4$  cm. 
Electromagnetic clusters with a small longitudinal extent are selected by requiring that 
the difference 
in the amounts of energy found in 
the second and fourth longitudinal layers of the electromagnetic 
calorimeter must be 
more than 40\% of the total cluster 
energy. 

Following this selection approximately 5500 (2000) $\pi^\circ$ candidates 
remain after the transverse momentum cut 
$p_{T,\pi}^* >2.5$ $(3.5)$ GeV. 
Using simulated events generated with the {\sc Lepto} 
and {\sc Ariadne} models 
the efficiency of the selection is estimated to be approximately  
45\%, and the contribution of
non-$\pi^{\circ}$ background is about 20\%.

The contamination of the sample from sources of high energy single photons 
other than  
$\pi^\circ$ decays, such as prompt photon production, is 
expected to be negligible since the rate of such processes is 
low\cite{prompt}. Using the {\sc Lepto} and {\sc Ariadne} models 
the total contamination due to 
the misidentification of electrons, and of  $\eta$-mesons decaying to two 
photons, is found to be less than  4\%. The measurements are corrected for 
this using 
the QCD-based models following the procedure outlined in section 4.3.
The contribution of the background from photoproduction processes was studied using the {\sc Phojet}\cite{phojet} model and 
found to be negligible.

\subsection{Correction Procedure and Systematic Uncertainties}
The results shown in this paper consist 
of two sets of spectra. First, the 
dependence of the $ep$ cross section for inclusive forward 
$\pi^\circ$-meson production on 
Bjorken-$x$, $Q^2$, $p_{T,\pi}^*$ and $x_\pi$ is studied. Measurements are 
then presented of the transverse energy flow in $\pi^\circ$-tagged 
events. The transverse energy is evaluated from energy deposits measured
in the LAr and SPACAL calorimeters, supplemented with tracking 
information from the CT, according to the prescription in \cite{HFS}.

The data are corrected using a bin-by-bin unfolding procedure using 
event samples generated with the {\sc Lepto} and {\sc Ariadne} models.   
The correction factors are obtained 
by taking the average of the correction factors estimated by these 
two models. 
The typical values of the correction factors obtained are 
approximately equal to 1.5. 

The following sources of systematic uncertainty are considered. The 
uncertainties from each source are added quadratically to form the total 
point-to-point systematic uncertainties on each 
of the measured distributions presented here: 
\begin{itemize}
\item The model dependence of the bin-by-bin acceptance corrections  
leads to systematic uncertainties of between 4\% and 11\% on 
both the $\pi^\circ$ cross section and transverse energy spectra.  
This source of uncertainty is calculated as half  
the difference of the correction factors derived from {\sc Lepto} and 
{\sc Ariadne}. The uncertainty resulting from the model dependence is 
largest at 
the lowest values of $x$ and for the highest values of  $p_{T,\pi}^*$ and 
$x_\pi$.
\item The systematic uncertainty due to the correction for 
QED radiative effects, calculated 
using {\sc Heracles}, is typically 
3\% for both the $\pi^0$ cross section and the transverse energy flow 
measurements. 
\item Variation of the $\pi^\circ$ selection and identification cuts 
within the 
resolution of the reconstructed quantities gives rise to an uncertainty 
of 5\% to 10\% 
in the measurements
of both the cross section and the transverse energy flow. 
\item
The uncertainty on the electromagnetic energy scale of 
the SPACAL (1\%) 
affects the reconstructed kinematics of the scattered positron. 
This results in uncertainties  
on the $\pi^\circ$ cross section and the transverse energy flow 
measurements of  
typically 9\% and 2\%, respectively.
\item
The uncertainty on the electromagnetic energy scale of the LAr (3\%) leads 
to  
an uncertainty on  the 
$\pi^\circ$ cross section of  5\% to 10\%, but has 
a negligible impact on 
the transverse energy spectra.
\item
The hadronic energy scale uncertainty on the LAr (4\%) 
gives rise to 
an uncertainty on the transverse energy flow measurements  of 4\%, but has 
negligible impact 
on the $\pi^\circ$ cross sections.
\item
The uncertainty on the hadronic energy scale of the SPACAL (7\%) results
in uncertainties on the $\pi^\circ$ cross section and the transverse 
energy flow
measurements of typically 2\% and 5\%, respectively.
\item
The uncertainty on the polar angle measurement of the scattered electron
(0.5 mrad) has a small influence (below 2\%)
on the $\pi^\circ$ cross section and the transverse energy flow spectra.
\item
The uncertainty on the determination of the trigger efficiency leads
to a 5\% uncertainty on the $\pi^\circ$ cross section measurements, but 
has 
a negligible effect on the transverse energy flow spectra.
\item
The uncertainty on the luminosity measurement leads to a 1.5\% uncertainty 
on the $\pi^\circ$ cross section measurement, but has no effect on the 
transverse energy distributions. 
\end{itemize}

\section{Results}

\subsection{Inclusive Forward \boldmath$\pi^\circ$ Cross Sections}
The inclusive $\pi^\circ$ cross section is measured differentially 
as a function of $Q^2$ and $x$, and as a function of $p_{T,\pi}^*$
and $x_{\pi}$, for $\pi^\circ$-mesons produced in the range  $p_{T,\pi}^* 
> $ 2.5 GeV, 5$^\circ$ $<\theta_{\pi} <$ 25$^\circ$ and $x_{\pi} >$ 0.01. 
The DIS phase space is restricted to the kinematic range 2 $< Q^2 < $ 70 
GeV$^2$ and 0.1 $< y < $ 0.6. The cross section data presented in this 
section are also given in Tables 1 and 2.

\begin{figure}[htb]
\begin{center}
\epsfig{file=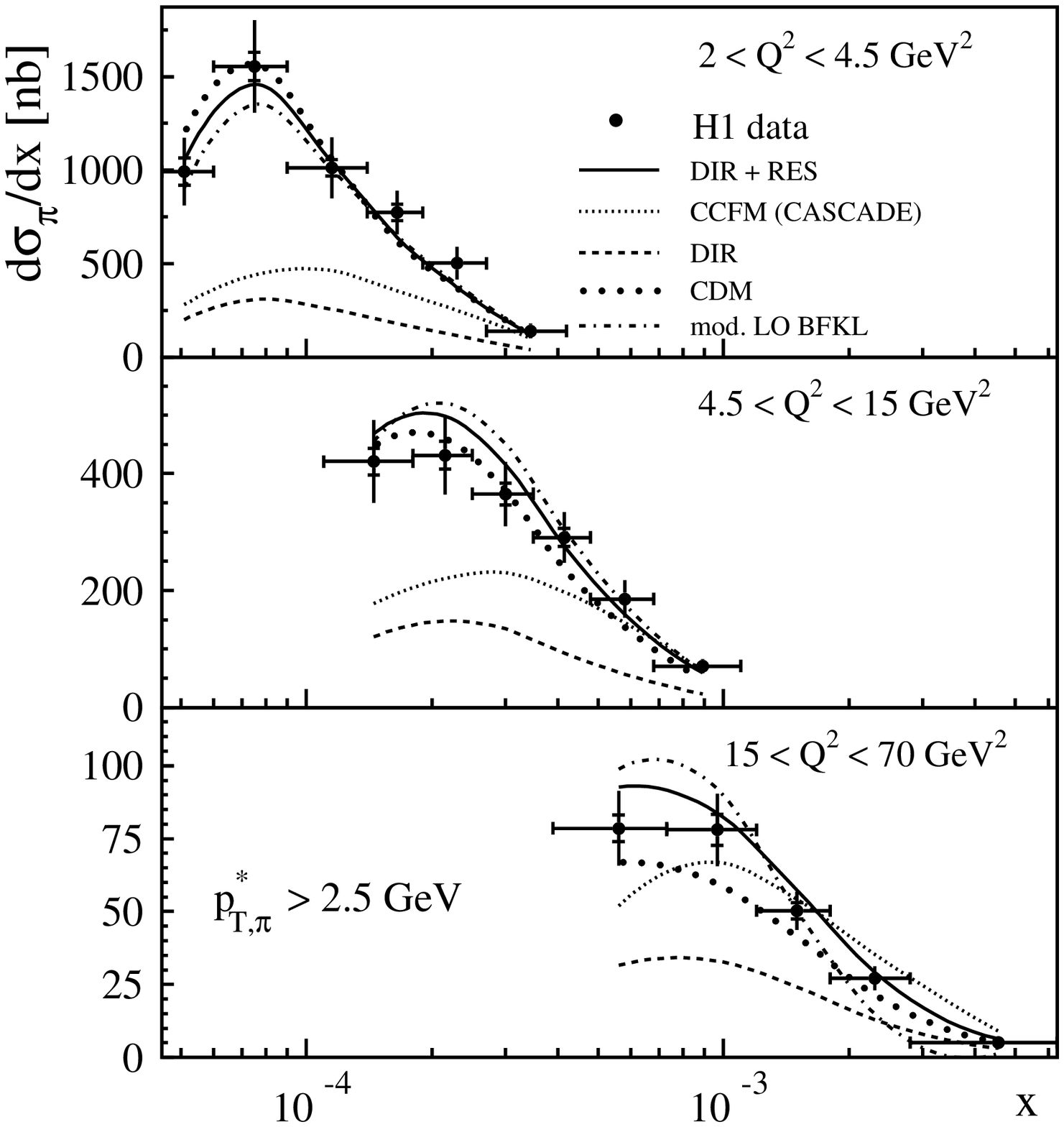,width=14.2cm,height=14.2cm}
\caption{The inclusive $ep$ cross section for forward 
$\pi^\circ$ mesons produced in the range
 $p_{T,\pi}^*>2.5$ GeV, $5^\circ < \theta_{\pi} < 25^\circ$ and 
$x_{\pi}=E_{\pi}/E_p >$0.01 as a function 
of Bjorken-$x$ in three intervals of $Q^2$. The DIS kinematic region is 
further specified by $0.1<y<0.6$.  
The inner error bars denote the statistical uncertainties and the outer 
error bars show the statistical and systematic uncertainties added 
quadratically. The predictions of five QCD-based 
models discussed in the text are shown.}
\label{fig1}
\end{center}
\end{figure}

The inclusive cross section $d\sigma_{\pi}/dx$ for $p_{T,\pi}^* > $ 2.5 GeV
as a function of Bjorken $x$ is shown in  Fig.~\ref{fig1} for three 
intervals of $Q^2$: 
$2 < Q^2 \leq 4.5$ GeV$^2$; $4.5 < Q^2  
\leq 15 $ GeV$^2$ 
and $15 < Q^2  < 70 $ GeV$^2$. The distributions rise with falling $x$ 
except at values of $x$ of around 
$10^{-4}$ in the lowest $Q^2$ region. This turnover is due to the 
limitations in 
the phase space imposed by the $\pi^\circ$ and DIS event selection cuts.

 The predictions of five QCD-based models are compared with the data. 
Calculations from 
{\sc Rapgap} which implement DGLAP 
evolution for proton 
structure only, labelled {\sc DIR}, fall substantially 
below the data. The 
disagreement becomes more pronounced at lower values of 
Bjorken-$x$. 
Calculations which assume virtual photon structure, marked {\sc DIR+RES}, 
describe the 
data well, although it is necessary to use rather large renormalisation
and factorisation scales, $\mu^2=Q^2 + 4p^2_T$, in order
to get a sufficiently large resolved photon component.
Using the same scale in {\sc Rapgap} a reasonable 
description of the azimuthal jet separation in a measurement of inclusive 
dijet production
at low $x$ in DIS\cite{dijet} is obtained. Forward jet data are also well 
described by DGLAP-based calculations which 
include virtual photon structure although different measurements tend to
prefer different choices of scales\cite{zeusfj,fjet2,smallX},
$\mu^2=Q^2 + 4p^2_T$ or $\mu^2=Q^2 + p^2_T$.   
Using the latter value would result in a reduction of about $30$\% in the 
normalisation of the predicted forward 
$\pi^\circ$ distributions measured in this paper, with little change to  
the shape.
Predictions based on the CCFM  equation, labelled {\sc CCFM (CASCADE)},  
agree at the highest values of $Q^2$. However, they fall below 
the data at low values of $x$ and $Q^2$. The rate of forward jet events 
predicted by {\sc Cascade} agrees with the data\cite{smallX}. This 
observation is 
not in contradiction with the forward  $\pi^\circ$ data since 
the discrepancies  observed here arise at the lowest values of $x$, which 
are not covered by the forward jet measurement. 
Differences between {\sc Rapgap} and {\sc Cascade} in the overall 
description of forward jet and particle 
production may be related to differences 
in the modelling of the partonic structure of forward jets. {\sc Cascade} 
mostly produces gluon-induced jets, while {\sc Rapgap} has a substantial 
contribution of quark-induced jets in 
the forward region. Fewer high momentum particles are produced in 
gluon-induced jets than in quark-induced jets~\cite{pjets}.
Predictions of the CDM give a reasonable description of the data.
 Analytical calculations using a modified 
BFKL equation, labelled 
  mod. {\sc LO BFKL}, describe the data well in the lower $Q^2$ region 
although they have a tendency 
to exceed the data at the lowest values of Bjorken-$x$ in 
the highest $Q^2$ 
intervals. 

\begin{figure}[htb]
\begin{center}
\epsfig{file=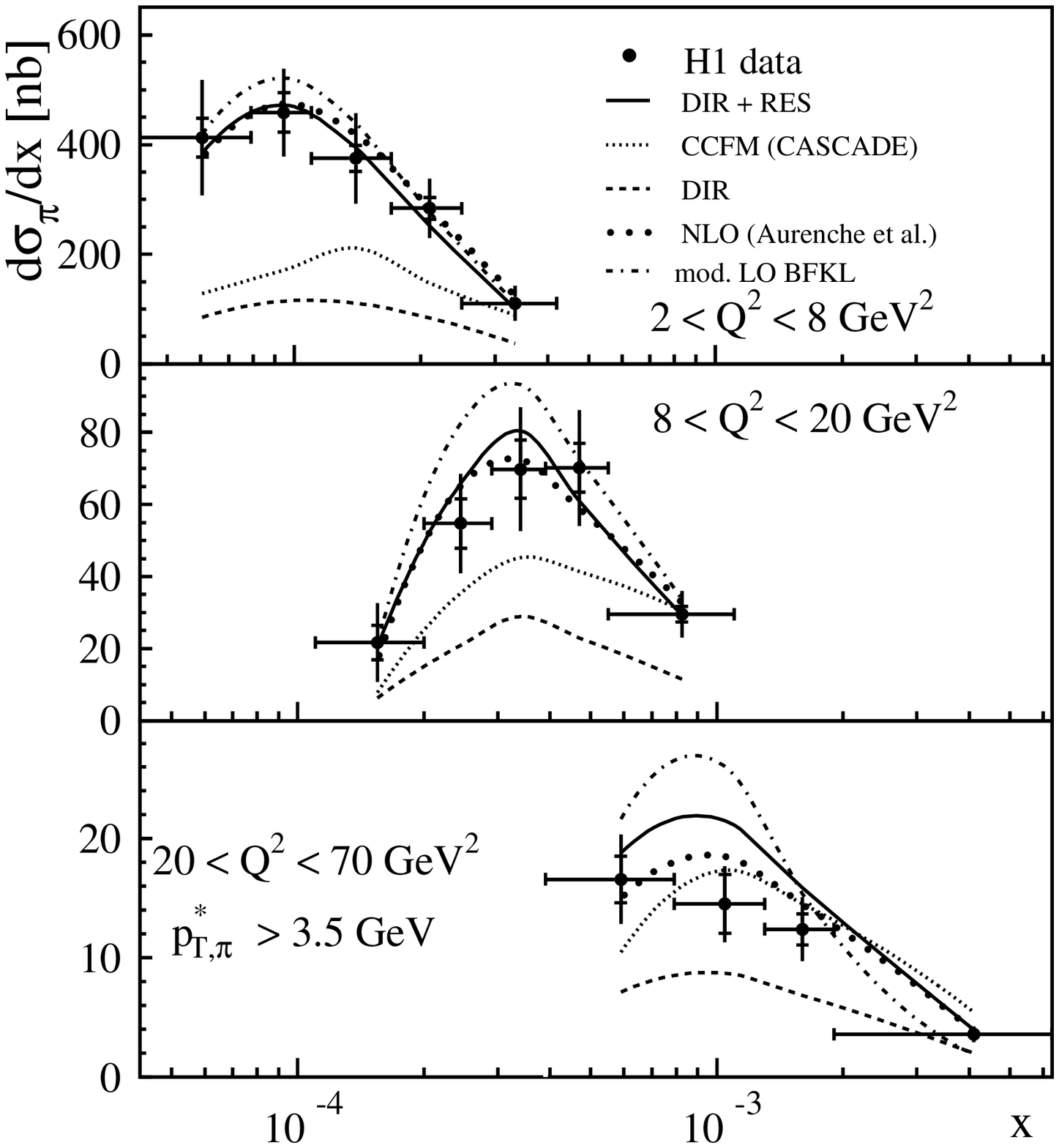,width=14.2cm,height=14.2cm}
\caption{
The inclusive $ep$ cross section for forward                           
$\pi^\circ$ mesons produced in the range                                        
 $p_{T,\pi}^*>3.5$ GeV, $5^\circ < \theta_{\pi} < 25^\circ$ and                 
$x_{\pi}=E_{\pi}/E_p >$0.01 as a function                                       
of Bjorken-$x$ in three intervals of $Q^2$. The DIS kinematic region is             
further specified by $0.1<y<0.6$.                                               
The inner error bars denote the statistical uncertainties and the outer         
error bars show the statistical and systematic uncertainties added              
quadratically. The predictions of five QCD-based                                
models discussed in the text are shown.}
\label{fig2}
\end{center}
\end{figure}

Fig.~\ref{fig2} shows the differential cross section $d\sigma_{\pi}/dx$
for transverse momenta $p_{T,\pi}^* > 3.5$ GeV,
 in three  intervals
of  $Q^2$: $2 < Q^2 \leq 8$ GeV$^2$; $8 < Q^2  \leq 20$ GeV$^2$
and $20 < Q^2  < 70$ GeV$^2$. Compared with the spectra in 
Fig.~\ref{fig1}, the measured differential                   
cross sections are lower by factors of between two and four. The QCD-based 
models provide a similar quality of description to that given in 
Fig.~\ref{fig1}. Calculations based on the modified BFKL equation 
clearly exceed 
the data in the two highest $Q^2$ intervals. 
The prediction of the NLO calculation\cite{aurenche} is also shown and  
describes the data well. Predictions of the CDM (not shown) overestimate 
the data at the lowest values of $Q^2$ and $x$.

The cross section $d\sigma_{\pi}/dp_{T,\pi}^*$ is shown as 
a function of $p_{T,\pi}^*$ in 
Fig.~\ref{fig3} in the same $Q^2$ intervals as in Fig.~\ref{fig1}. 
The data 
fall steeply with increasing $p_{T,\pi}^*$ and the shapes of the 
distributions vary only slightly with increasing $Q^2$. 
The calculations implementing resolved virtual photons describe the data 
well. 
The predictions of the model including only direct processes fall  
below the data everywhere 
although they come  nearer to the data 
as $Q^2$ increases. The CCFM-based calculations fail in the 
lowest $Q^2$ interval but give a reasonable description of the highest 
$Q^2$ interval. 
The CDM predicts spectra which are somewhat harder than 
the data. 

In Fig.~\ref{fig4} the cross section 
$d\sigma_{\pi}/dx_\pi$ is shown as a function of  
$x_\pi$ in the same three intervals of $Q^2$ as used in Fig.~\ref{fig1}.  
In Fig.~\ref{fig5} the differential cross section is presented in 
three intervals of Bjorken-$x$: $4.2\cdot 10^{-5}< x \leq 2 \cdot
10^{-4}; 2\cdot 10^{-4}< x \leq 10^{-3}$ and $10^{-3}< x < 6.3 \cdot 
10^{-3}$. The cross section falls as $x_\pi$ increases. There is no 
strong dependence of the shapes of 
the distributions on Bjorken-$x$ or $Q^2$. The 
resolved photon approach, the CDM, and the BFKL calculations
describe the spectra well.
The CCFM 
implementation describes the data only in the highest intervals of $Q^2$ 
and $x$. The direct photon calculations fall below the data in all of the 
spectra but approach them as $Q^2$ increases.\\

\subsection{Transverse Energy Flow}
The transverse energy flow, 
defined as the mean transverse energy per 
event per unit of pseudorapidity difference 
in the hadronic centre of mass system, 
$\frac{1}{N}d E_T^*/d(\eta^*-\eta^*_\pi)$, in events 
containing at least one 
forward $\pi^\circ$ is presented in                  
Fig.~\ref{fig6}. 
Here $N$ is the total number of events and $E_{T}^*$ is 
the sum of the transverse energies of each particle $i$: $E_{T}^*=\Sigma_i 
E_{Ti}^*$. The transverse 
energy $E_{Ti}^*$ of a particle $i$ with energy 
$E_i^*$ and polar angle $\theta_i^*$ is defined as $E_{Ti}^*=E_i^* 
\sin\theta^*_i$. The pseudorapidity $\eta^*$ is defined as 
$-\ln\tan(\theta^*/2)$. The energy flow, which includes the contribution 
from the forward $\pi^\circ$, is plotted as a function of the difference 
in 
pseudorapidity $\eta^*-\eta_\pi^*$ from the selected forward 
$\pi^\circ$. In events containing more than one forward $\pi^\circ$, 
the candidate with the largest transverse momentum is chosen.
The spectra are presented in three intervals of the $\pi^\circ$ pseudorapidity
ranging from close (Fig.~7a) to far (Fig.~7c) from the proton direction.

The spectra show a large increase of transverse energy production  
in the region associated with the $\pi^\circ$. This can be understood as 
being due to the energy associated with the jet which contains 
the leading $\pi^\circ$.
 A broad distribution of lower transverse energy flow  
 in the current region reveals the range over which the transverse 
momentum of the jet is compensated.

\begin{figure}[htb]
\begin{center}
\epsfig{file=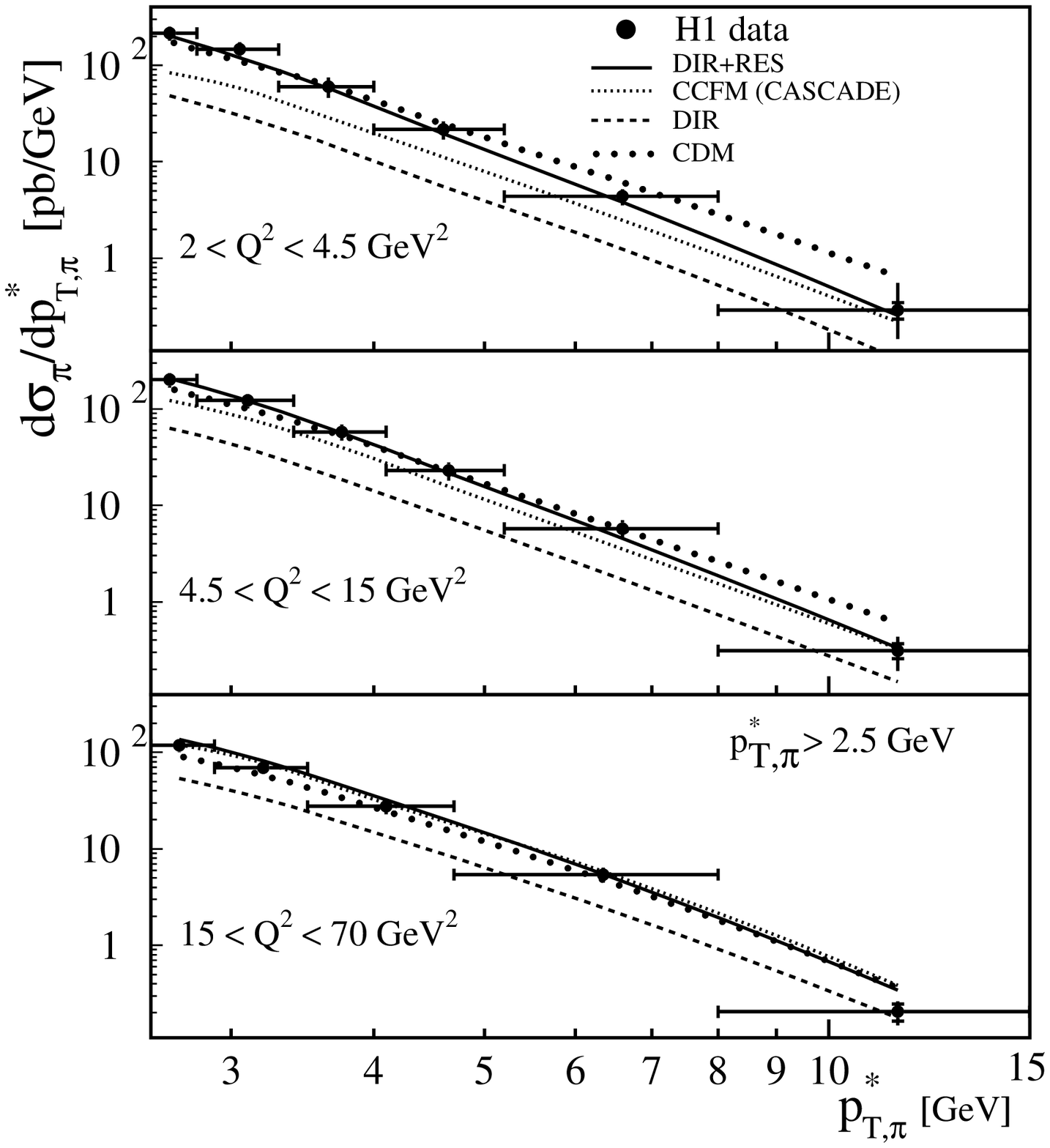,width=14.2cm,height=14.2cm}
\caption{
The inclusive $ep$ cross section for forward                           
$\pi^\circ$ mesons produced in the range                                        
 $p_{T,\pi}^*>2.5$ GeV, $5^\circ < \theta_{\pi} < 25^\circ$ and                 
$x_{\pi}=E_{\pi}/E_p >$0.01 as a function                                       
of $p_{T,\pi}$ in three intervals of $Q^2$. The DIS kinematic region is             
further specified by $0.1<y<0.6$.                                               
The inner error bars denote the statistical uncertainties and the outer         
error bars show the statistical and systematic uncertainties added              
quadratically. The predictions of four QCD-based                                
models discussed in the text are shown.}
\label{fig3}
\end{center}
\end{figure}

\begin{figure}[htb]
\begin{center}
\epsfig{file=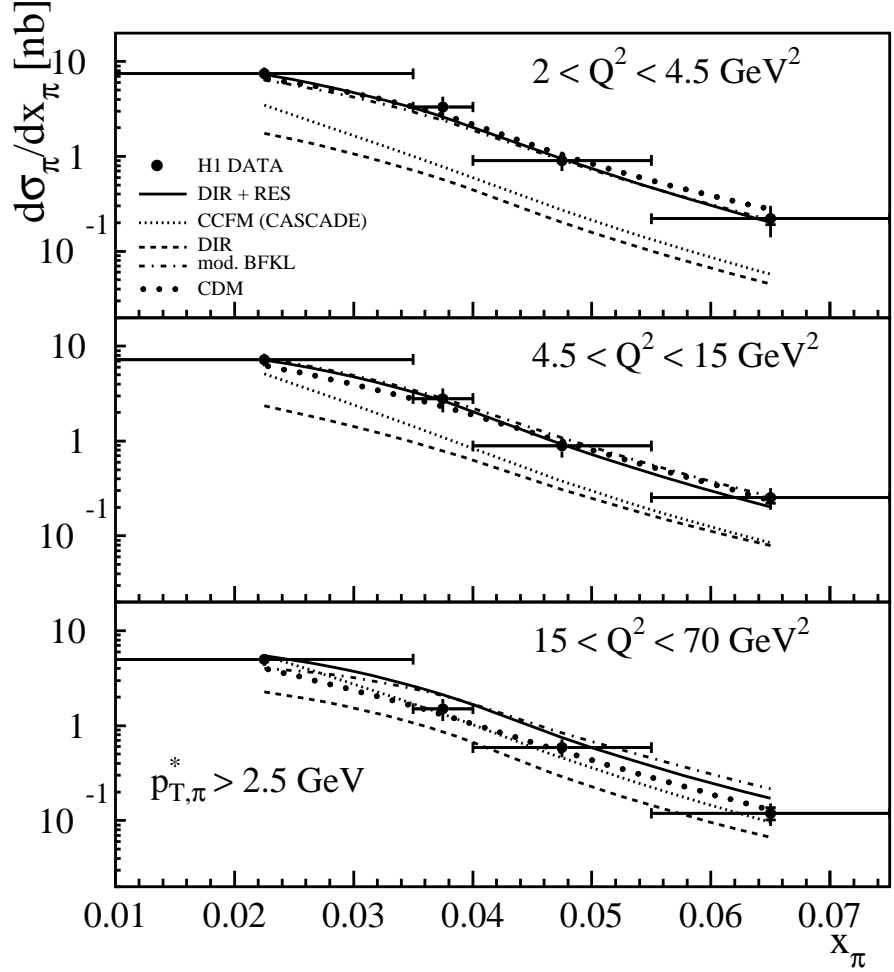,width=14.2cm,height=14.2cm}
\caption{
The inclusive $ep$ cross section for forward                           
$\pi^\circ$ mesons produced in the range                                        
 $p_{T,\pi}^*>2.5$ GeV, $5^\circ < \theta_{\pi} < 25^\circ$ as a function                                       
of $x_\pi$ in three intervals of $Q^2$. The DIS kinematic region is             
further specified by $0.1<y<0.6$.                                               
The inner error bars denote the statistical uncertainties and the outer         
error bars show the statistical and systematic uncertainties added              
quadratically. The predictions of five QCD-based                                
models discussed in the text are shown.}
\label{fig4}
\end{center}
\end{figure}

\begin{figure}[htb]
\begin{center}
\epsfig{file=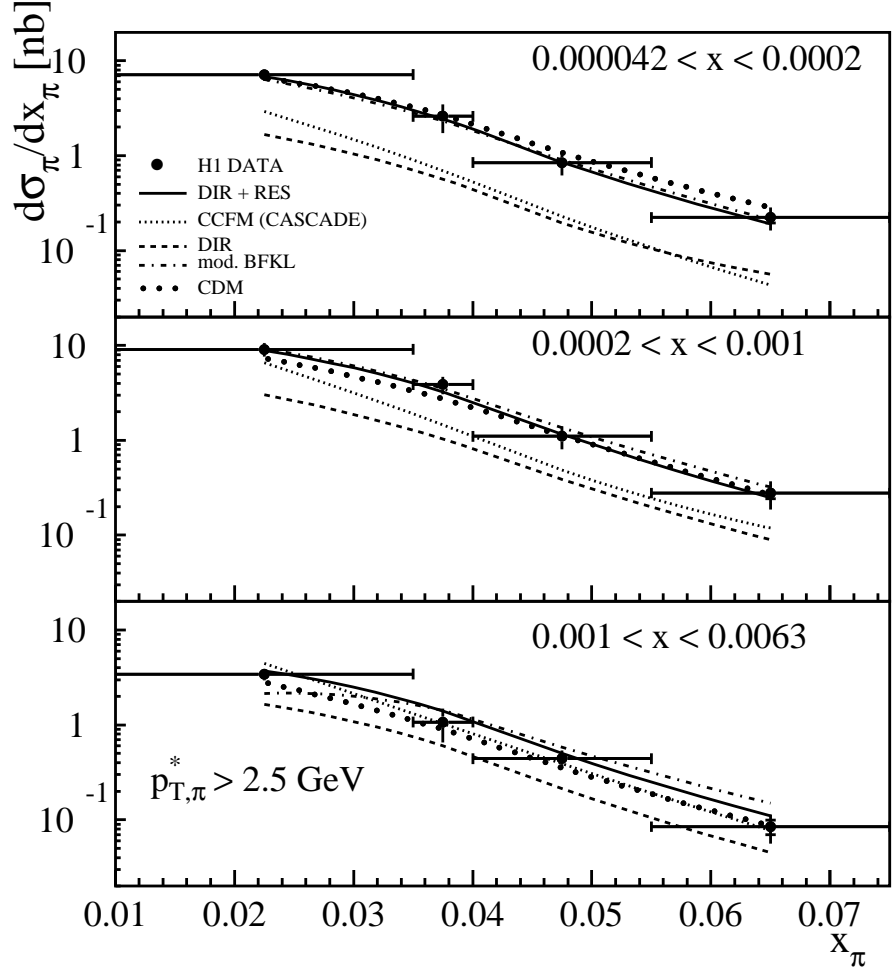,width=14.2cm,height=14.2cm}
\caption{
The inclusive $ep$ cross section for forward                           
$\pi^\circ$ mesons produced in the range                                        
 $p_{T,\pi}^*>2.5$ GeV and $5^\circ < \theta_{\pi} < 25^\circ$ as a 
function                                       
of $x_\pi$ in three intervals of Bjorken-$x$. The DIS kinematic region is             
further specified by $2 < Q^2< 70$ GeV$^2$ and $0.1<y<0.6$.                                              
The inner error bars denote the statistical uncertainties and the outer         
error bars show the statistical and systematic uncertainties added              
quadratically. The predictions of five QCD-based                                
models discussed in the text are shown.}
\label{fig5}
\end{center}
\end{figure}

\begin{figure}[htb]
\begin{center}
\epsfig{file=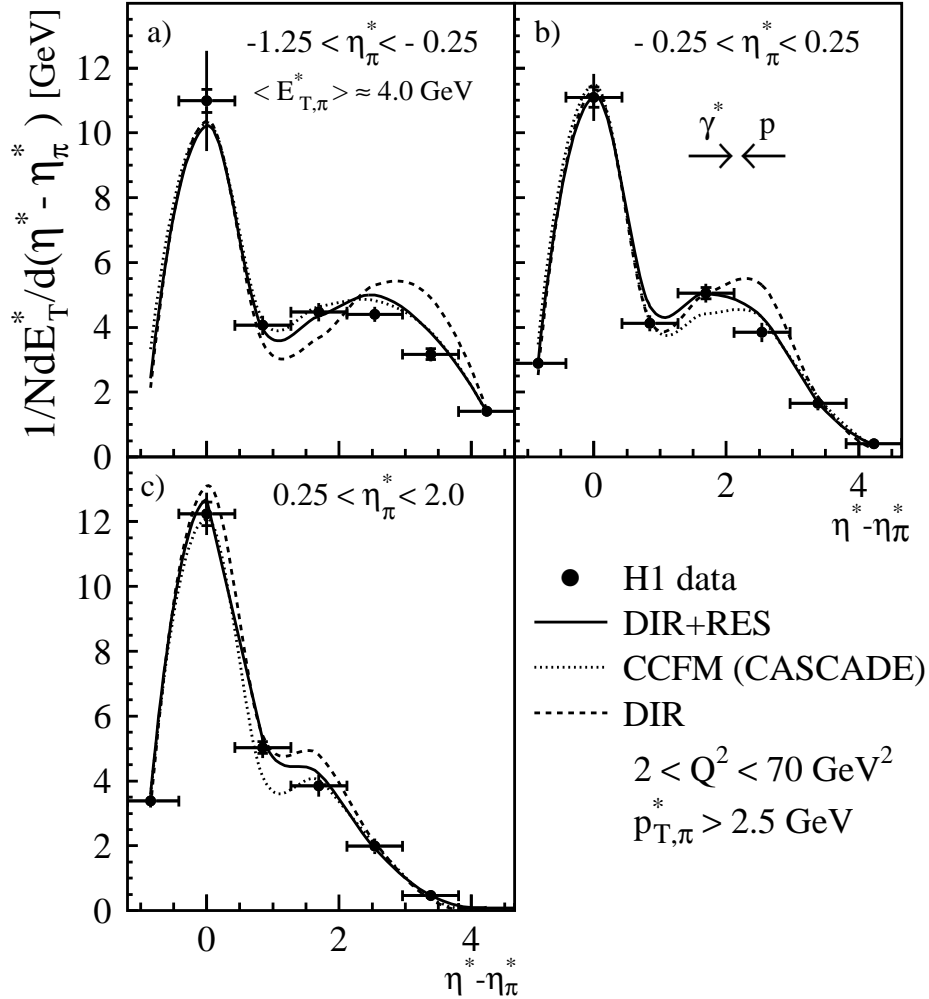,width=14.2cm,height=14.2cm}
\caption{Distributions of transverse energy flow in events containing a  
 forward $\pi^\circ$ produced in the range $p_{T,\pi}^*>2.5$ GeV, $5^\circ 
< \theta_{\pi} < 25^\circ$ and                  
$x_{\pi}=E_{\pi}/E_p >$ 0.01. 
The transverse energy flow is presented 
as a function of the distance in pseudorapidity from the selected 
forward $\pi^\circ$ for various ranges in the $\pi^\circ$ pseudorapidity. 
The DIS kinematic region is specified by $2 < Q^2 < 70$ GeV$^2$ and $0.1 
< y < 0.6$. The inner error bars denote the statistical uncertainties and 
the outer error bars show the statistical and systematic uncertainties 
added              
quadratically. The predictions of three QCD-based                                
models discussed in the text are shown. }
\label{fig6}
\end{center}
\end{figure}

The QCD-based models all describe the transverse energy 
flow in the vicinity of the $\pi^\circ$ but give 
different predictions in the current region.
Calculations with resolved photon processes tend to agree best with
the data. 
The CCFM approach provides a reasonable description of the data. This 
model predicts a strong 
compensation of the $\pi^\circ$ transverse momentum in the 
rapidity region  between the forward particle and the proton remnant 
system which is, however,  not covered in the measurements presented here.
The direct photon model shows a peak at larger 
values of pseudorapidity difference $\eta^*-\eta_\pi^*$ than is observed 
in the data. This effect becomes less pronounced with 
increasing $\pi^\circ$ pseudorapidity as the forward  $\pi^\circ$ may
enter the current jet region.  
The differences between the models can be qualitatively understood as 
a consequence of the ordering or otherwise of the $k_T$ in the parton 
cascades. The $k_T$ ordering
in the direct photon model forces the compensation of the transverse energy   
in the current region while models without this requirement allow compensation 
close to the $\pi^\circ$-meson.
The CDM (not shown)  predicts too much transverse energy in the vicinity 
of the $\pi^\circ$. This is a consequence of the overly hard 
$\pi^\circ$ transverse momentum distribution predicted by the CDM 
and shown earlier in Fig.~\ref{fig3}. 

\begin{figure}[htb]
\begin{center}
\epsfig{file=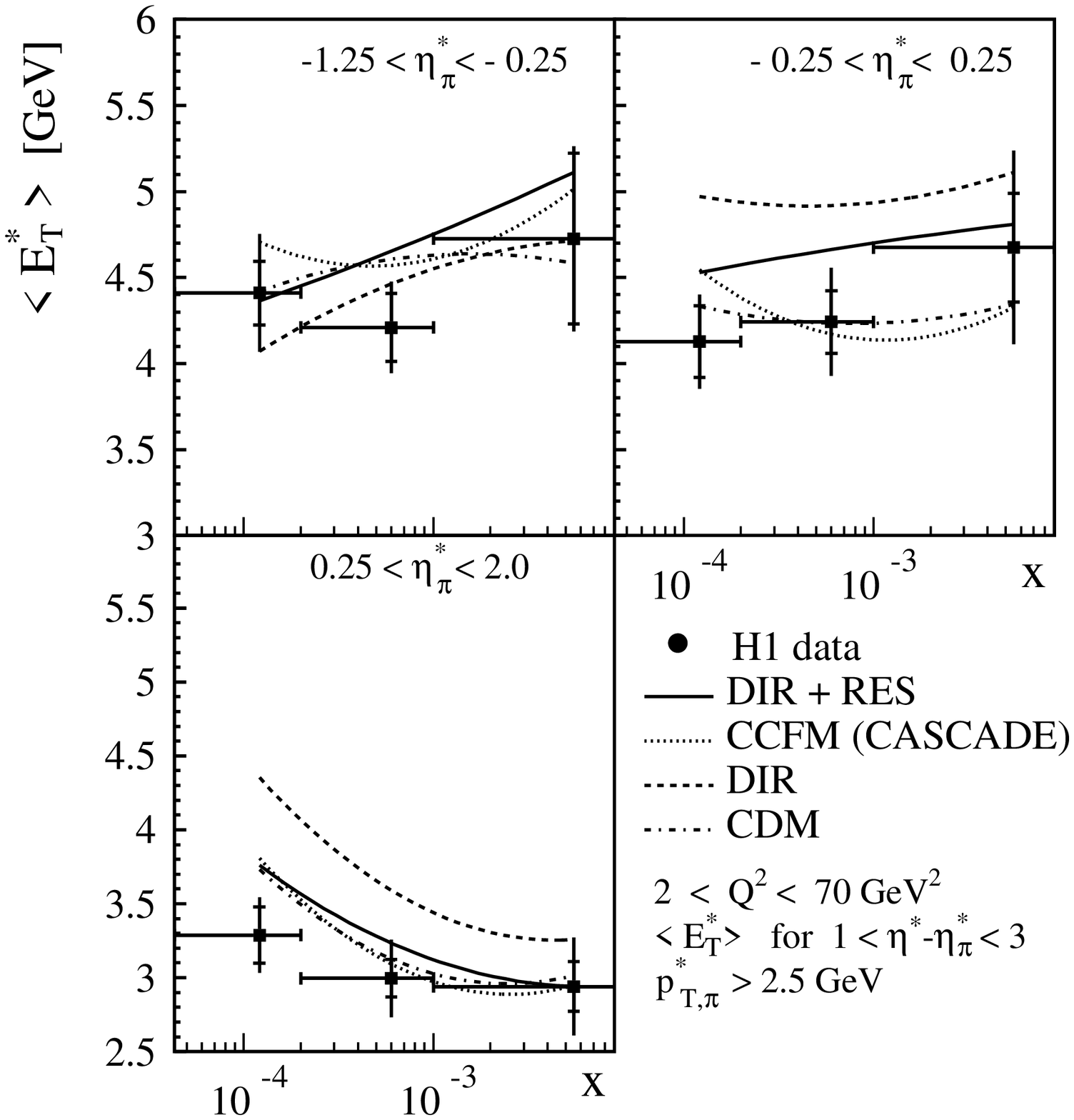,width=14.2cm,height=14.2cm}
\caption{The mean transverse energy in events containing a forward 
$\pi^\circ$ produced in the range $p_{T,\pi}^*>2.5$ GeV, $5^\circ      
< \theta_{\pi} < 25^\circ$ and  $x_{\pi}=E_{\pi}/E_p >$ 0.01. The 
transverse energy is measured over the region 
$1.0 < \eta^* - \eta^*_\pi < 3.0$  as a function of 
Bjorken-$x$ for three intervals of $\pi^\circ$ pseduorapidity. 
The DIS kinematic region is further specified by $2 < Q^2 < 70$ GeV$^2$ and $0.1 
< y < 0.6$. The inner error bars denote the statistical uncertainties and 
the outer error bars show the statistical and systematic 
uncertainties added quadratically. The predictions of four QCD-based 
models discussed in the text are shown. }        
\label{fig7}
\end{center}
\end{figure}

Transverse energy flow in the region away from the 
forward $\pi^\circ$ and into the 
current region is further studied as shown in
Fig.~\ref{fig7}. The mean transverse energy over the region 
$1.0 < \eta^* - \eta^*_\pi < 3.0$ is 
plotted as a function of Bjorken-$x$ for the same 
three intervals of the $\pi^\circ$ 
pseudorapidity as in the previous figure.
The data 
show no significant dependence on Bjorken-$x$  
although there is a tendency for the mean transverse energy to fall as  
pseudorapidity of $\pi^\circ$ approaches the current region. 
With the exception of the direct photon model, all of the QCD-based models 
give a reasonable description of the data. 
The direct photon  prescription is only able to describe the data 
in the pseudorapidity region closest to the proton remnant. 

\section{Summary} 
Measurements are presented of $ep$ interactions containing high 
transverse momentum  
forward going $\pi^\circ$-mesons in the deep-inelastic 
scattering regime of $0.1 < y< 
0.6$, $2 < Q^2 < 70$~GeV$^2$ and $4 \cdot 10^{-5} < x <6 \cdot 10^{-3}$.  
The inclusive $\pi^\circ$ cross section is measured differentially 
as a function of 
$Q^2$, Bjorken-$x$ and $\pi^\circ$ transverse momentum and  energy for particles with 
$p_{T,\pi}^*> 2.5$ GeV, $5^\circ < \theta_\pi < 25^\circ$ and $x_\pi=\frac{E\pi}{E_{p}}>0.01$. 
The transverse energy flow relative to the direction of the forward  
$\pi^\circ$ and the mean transverse energy in the vicinity of the 
forward $\pi^\circ$ are also measured.  

The measurements presented are sensitive to the dynamics of parton evolution. 
Several different QCD-based approaches are confronted with the data. 
An approach implementing DGLAP evolution of proton structure 
underestimates the $\pi^\circ$ cross section 
at low values of Bjorken-$x$ and $Q^2$ and overestimates the range at which the 
$\pi^\circ$ transverse momentum is compensated. Calculations implementing 
virtual photon  structure provide the best description, albeit with a 
preferred choice of  renormalisation 
and factorisation scale which is inconsistent with that required by  
other measurements of the hadronic final state in the forward region. 
Predictions based on CCFM evolution fail to describe the $\pi^\circ$ 
cross section at the 
lowest values of Bjorken-$x$ but give 
a fair description of the transverse energy distributions. 
The Colour Dipole Model gives a reasonable description of the $\pi^\circ$ 
cross section but predicts a transverse momentum distribution of 
$\pi^\circ$-mesons which is significantly harder than is observed in the 
data. Calculations using next-to-leading 
order QCD matrix elements convoluted  with $\pi^\circ$ fragmentation 
functions describe  the Bjorken-$x$ dependence of the forward $\pi^\circ$ 
cross section well. Predictions made using a modified BFKL equation 
describe the forward $\pi^\circ$ cross section at the lowest $Q^2$ values but 
exceed the data at the highest $Q^2$ values.

\section*{Acknowledgements}
We would like to  thank  Alan Martin, John Outhwaite and the late Jan 
Kwieci\'nski for
making available the program to calculate forward jet and particle 
cross sections using the modified LO BFKL model. We are grateful to  
Patrik Aurenche, Michel Fontannaz, Rohini Godbole and Rahul Basu for 
providing their NLO calculations. We also thank Gunnar 
Ingelman 
for useful discussions. We are grateful to the HERA machine group whose 
outstanding efforts have made this experiment 
possible. We thank the engineers and
technicians for their work in constructing and now maintaining the H1 
detector, our
funding agencies for financial support, the DESY technical staff for continual
assistance and the DESY directorate for support and for the hospitality 
which they extend to the non-DESY members of the collaboration.

\newpage

\setlength{\textwidth}{14cm} \setlength{\textheight}{22cm}
\linespread{1.2}
\begin{table}[tp]
\begin{center}
\begin{tabular}{|c|c|c|c|}

\hline
$x\cdot10^4$&($\frac{d\sigma_\pi}{dx})^{\pm{stat}}_{\pm{tot}}$ 
(nb)&$x\cdot10^4$&($\frac{d\sigma_\pi}{dx})^{\pm{stat}}_{\pm{tot}}$ (nb)\\
\hline\hline
\multicolumn{1}{|c}{\textbf{$2 < Q^2 < 4.5\hspace{3pt} \rm GeV^2$}} &
\multicolumn{1}{c|}{\textbf{$p^*_{T,\pi} > 2.5\hspace{3pt}\rm GeV $}} &
\multicolumn{1}{|c}{\textbf{$2< Q^2 < 8\hspace{3pt} \rm GeV^2$}} &
\multicolumn{1}{c|}{\textbf{$p^*_{T,\pi} > 3.5\hspace{3pt}\rm GeV $}}\\
\hline
0.42---0.60&\textbf{$990^{\pm{70}}_{\pm{180}}$}&0.42---0.79&\textbf{$410^{\pm{40}}_{\pm{110}}$}\\
0.60---0.90&\textbf{$1550^{\pm{80}}_{\pm{250}}$}&0.79---1.1&\textbf{$459^{\pm{36}}_{\pm{80}}$}\\
0.90---1.4&\textbf{$1010^{\pm{50}}_{\pm{160}}$}&1.1---1.7&\textbf{$375^{\pm{24}}_{\pm{82}}$}\\
1.4---1.9&\textbf{$770^{\pm{40}}_{\pm{120}}$}&1.7---2.5&\textbf{$284^{\pm{20}}_{\pm{54}}$}\\
1.9---2.7&\textbf{$503^{\pm{28}}_{\pm{88}}$}&2.5---4.2&\textbf{$110^{\pm{8}}_{\pm{32}}$}\\
2.7---4.2&\textbf{$141^{\pm{10}}_{\pm{40}}$}&{}&{}\\
\hline\hline
\multicolumn{1}{|c}{\textbf{$4.5 < Q^2 < 15\hspace{3pt} \rm GeV^2$}} &
\multicolumn{1}{c|}{\textbf{$p^*_{T,\pi} > 2.5\hspace{3pt}\rm GeV $}} &
\multicolumn{1}{|c}{\textbf{$8< Q^2 < 20\hspace{3pt} \rm GeV^2$}} &
\multicolumn{1}{c|}{\textbf{$p^*_{T,\pi} > 3.5\hspace{3pt}\rm GeV $}}\\
\hline
1.1---1.8&\textbf{$421^{\pm{23}}_{\pm{71}}$}&1.1---2.0&\textbf{$22^{\pm{5}}_{\pm{11}}$}\\
1.8---2.5&\textbf{$432^{\pm{24}}_{\pm{67}}$}&2.0---2.9&\textbf{$55^{\pm{7}}_{\pm{14}}$}\\
2.5---3.5&\textbf{$365^{\pm{19}}_{\pm{55}}$}&2.9---3.9&\textbf{$70^{\pm{8}}_{\pm{17}}$}\\
3.5---4.8&\textbf{$291^{\pm{15}}_{\pm{43}}$}&3.9---5.5&\textbf{$70^{\pm{7}}_{\pm{16}}$}\\
4.8---6.8&\textbf{$185^{\pm{10}}_{\pm{32}}$}&5.5---11.0&\textbf{$29.5^{\pm{2.1}}_{\pm{6.4}}$}\\
6.8---11.0&\textbf{$70^{\pm{4}}_{\pm{12}}$}&{}&{}\\
\hline\hline
\multicolumn{1}{|c}{\textbf{$15 < Q^2 < 70\hspace{3pt} \rm GeV^2$}} &
\multicolumn{1}{c|}{\textbf{$p^*_{T,\pi} > 2.5\hspace{3pt}\rm GeV $}} &
\multicolumn{1}{|c}{\textbf{$20< Q^2 < 70\hspace{3pt} \rm GeV^2$}} &
\multicolumn{1}{c|}{\textbf{$p^*_{T,\pi} > 3.5\hspace{3pt}\rm GeV $}}\\
\hline
3.9---7.3&\textbf{$79^{\pm{5}}_{\pm{13}}$}&3.9---7.9&\textbf{$16.6^{\pm{1.9}}_{\pm{3.7}}$}\\
7.3---12.0&\textbf{$78^{\pm{5}}_{\pm{12}}$}&7.9---13.0&\textbf{$14.5^{\pm{2.4}}_{\pm{3.2}}$}\\
12.0---18.0&\textbf{$50.3^{\pm{2.8}}_{\pm{6.5}}$}&13.0---19.0&\textbf{$12.4^{\pm{1.3}}_{\pm{2.7}}$}\\
18.0---28.0&\textbf{$27.1^{\pm{1.5}}_{\pm{4.2}}$}&19.0---63.0&\textbf{$3.60^{\pm{0.27}}_{\pm{0.65}}$}\\
28.0---63.0&\textbf{$5.1^{\pm{0.4}}_{\pm{1.1}}$}&{}&{}\\
\hline
\end{tabular}
\\
\vspace{5pt}

\end{center}
\caption[]{The inclusive $\pi^\circ$-meson cross
sections as shown in Figs.~2 and 3, presented 
together with the statistical and total uncertainties.}

\end{table}

\setlength{\textwidth}{14cm} \setlength{\textheight}{22cm}
\linespread{1.2}
\begin{table}
\begin{center}
\begin{tabular}{|c|c|c|c|c|c|}
\hline
$p^*_{T,\pi}$&($\frac{d\sigma_\pi}{dp^*_{T,\pi}})^{\pm{stat}}_{\pm{tot}}$ 
($\frac{\rm pb}{\rm 
GeV}$)&$p^*_{T,\pi}$&($\frac{d\sigma_\pi}{dp^*_{T,\pi}})^{\pm{stat}}_{\pm{tot}}$ 
($\frac{\rm pb}{\rm 
GeV}$)&$p^*_{T,\pi}$&($\frac{d\sigma_\pi}{dp^*_{T,\pi}})^{\pm{stat}}_{\pm{tot}}$ 
($\frac{\rm pb}{\rm GeV}$)\\
\hline\hline
\multicolumn{2}{|c|}{$ 2 < Q^2 < 4.5 \hspace{3pt} \rm GeV^2$}&
\multicolumn{2}{|c|}{$ 4.5 < Q^2 < 15 \hspace{3pt} \rm GeV^2$}&
\multicolumn{2}{|c|}{$ 15 < Q^2 < 70 \hspace{3pt} \rm GeV^2$}\\
\hline
2.5---2.8&\textbf{$215^{\pm{9}}_{\pm{36}}$}&2.5---2.8&\textbf{$200^{\pm{8}}_{\pm{34}}$}&2.5---2.9&\textbf{$118^{\pm{5}}_{\pm{18}}$}\\
2.8---3.3&\textbf{$146^{\pm{6}}_{\pm{27}}$}&2.8---3.4&\textbf{$122^{\pm{5}}_{\pm{21}}$}&2.9---3.5&\textbf{$69.1^{\pm{3.3}}_{\pm{8.8}}$}\\
3.3---4.0&\textbf{$60.0^{\pm{3}}_{\pm{14}}$}&3.4---4.1&\textbf{$58^{\pm{3}}_{\pm{11}}$}&3.5---4.7&\textbf{$27.7^{\pm{2.0}}_{\pm{4.7}}$}\\
4.0---5.2&\textbf{$21.7^{\pm{1.4}}_{\pm{4.7}}$}&4.1---5.2&\textbf{$23.0^{\pm{1.5}}_{\pm{4.9}}$}&4.7---8.0&\textbf{$5.41^{\pm{0.38}}_{\pm{0.96}}$}\\
5.2---8.0&\textbf{$4.37^{\pm{0.38}}_{\pm{0.83}}$}&5.2---8.0&\textbf{$5.8^{\pm{0.4}}_{\pm{1.2}}$}&8.0---15.0&\textbf{$0.21^{\pm{0.04}}_{\pm{0.06}}$}\\
8.0---15.0&\textbf{$0.29^{\pm{0.06}}_{\pm{0.26}}$}&8.0---15.0&\textbf{$0.31^{\pm{0.06}}_{\pm{0.12}}$}&{}&{}\\
\hline
\end{tabular}
\end{center}
\caption[]{The inclusive $\pi^\circ$-meson cross
sections as shown in Fig.~4, presented
together with the statistical and total uncertainties.}
\end{table}

\setlength{\textwidth}{14cm} \setlength{\textheight}{22cm}
\linespread{1.2}
\begin{table}[tp]
\begin{center}
\begin{tabular}{|c|c|c|c|}
\hline
$x_\pi$&
\multicolumn{3}{|c|}{($\frac{d\sigma_\pi}{dx_\pi})^{\pm{stat}}_{\pm{tot}}$ 
(nb)}\\
\hline\hline
\multicolumn{4}{|c|}{\textbf{$p^*_{T,\pi} > 2.5\hspace{3pt}\rm GeV $}}\\
\hline
{}&$ 0.000042 < x < 0.0002$&$ 0.0002 < x < 0.001$&$ 0.001 < x < 0.0063$\\
\hline
0.01---0.035&\textbf{$7.2^{\pm{0.2}}_{\pm{1.0}}$}&\textbf{$9.1^{\pm{0.2}}_{\pm{1.5}}$}&\textbf{$3.44^{\pm{0.12}}_{\pm{0.48}}$}\\
0.035---0.04&\textbf{$2.60^{\pm{0.23}}_{\pm{0.88}}$}&\textbf{$3.87^{\pm{0.27}}_{\pm{0.81}}$}&\textbf{$1.07^{\pm{0.13}}_{\pm{0.41}}$}\\
0.04---0.055&\textbf{$0.85^{\pm{0.07}}_{\pm{0.23}}$}&\textbf{$1.10^{\pm{0.08}}_{\pm{0.30}}$}&\textbf{$0.44^{\pm{0.05}}_{\pm{0.10}}$}\\
0.055---0.075&\textbf{$0.22^{\pm{0.03}}_{\pm{0.06}}$}&\textbf{$0.28^{\pm{0.03}}_{\pm{0.09}}$}&\textbf{$0.08^{\pm{0.01}}_{\pm{0.03}}$}\\
\hline\hline
\multicolumn{4}{|c|}{\textbf{$p^*_{T,\pi} > 2.5\hspace{3pt}\rm GeV $}}\\
\hline
{}&$ 2 < Q^2 < 4.5\hspace{3pt} \rm GeV^2$&$ 4.5 < Q^2 < 15\hspace{3pt} \rm GeV^2$&$ 15 < Q^2 < 70\hspace{3pt} \rm GeV^2$\\
\hline
0.01---0.035&\textbf{$7.5^{\pm{0.2}}_{\pm{1.2}}$}&\textbf{$7.3^{\pm{0.2}}_{\pm{1.1}}$}&\textbf{$5.01^{\pm{0.16}}_{\pm{0.74}}$}\\
0.035---0.04&\textbf{$3.32^{\pm{0.28}}_{\pm{0.92}}$}&\textbf{$2.78^{\pm{0.23}}_{\pm{0.78}}$}&\textbf{$1.51^{\pm{0.14}}_{\pm{0.39}}$}\\
0.04---0.055&\textbf{$0.91^{\pm{0.07}}_{\pm{0.21}}$}&\textbf{$0.89^{\pm{0.07}}_{\pm{0.22}}$}&\textbf{$0.59^{\pm{0.05}}_{\pm{0.12}}$}\\
0.055---0.075&\textbf{$0.22^{\pm{0.03}}_{\pm{0.08}}$}&\textbf{$0.25^{\pm{0.03}}_{\pm{0.07}}$}&\textbf{$0.12^{\pm{0.02}}_{\pm{0.03}}$}\\
\hline
\end{tabular}
\end{center}
\caption[]{The inclusive $\pi^\circ$-meson cross
sections as shown in Figs.~5 and 6, presented
together with the statistical and total uncertainties.}
\end{table}

\newpage


\begin{thebibliography}{99}

\bibitem{et1} 
I.~Abt {\it et al.} [H1 Collaboration],
 Z.\ Phys.\ {\bf C 63} (1994) 377.
\bibitem{fjet1}
S.~Aid {\it et al.}  [H1 Collaboration],
Phys.\ Lett.\ {\bf B 356} (1995) 118
[hep-ex/9506012].
\bibitem{fjet2}
C.~Adloff {\it et al.}  [H1 Collaboration],
Nucl.\ Phys.\  {\bf B 538} (1999) 3
[hep-ex/9809028].
\bibitem{zeusfj}
J.~Breitweg {\it et al.}  [ZEUS Collaboration],
Eur.\ Phys.\ J.\ {\bf C 6} (1999) 239 
[hep-ex/9805016]; \\
J.~Breitweg {\it et al.}  [ZEUS Collaboration],
Phys.\ Lett.\ {\bf B 474} (2000) 223
[hep-ex/9910043].
\bibitem{DGLAP}
V.~N.~Gribov and L.~N.~Lipatov,
Sov.\ J.\ Nucl.\ Phys.\  {\bf 15} (1972) 438 and 675;\\
L.~N.~Lipatov,
 Sov.\ J.\ Nucl.\ Phys.\  {\bf 20} (1975) 94;\\
%
G.~Altarelli and G.~Parisi,
Nucl.\ Phys.\ {\bf B 126} (1977) 298;\\
%
Y.~L.~Dokshitzer,
Sov.\ Phys.\ JETP {\bf 46} (1977) 641.
\bibitem{BFKL}
E.~A.~Kuraev, L.~N.~Lipatov and V.~S.~Fadin,
Sov.\ Phys.\ JETP {\bf 45} (1977) 199;\\
%
I.~I.~Balitsky and L.~N.~Lipatov,
Sov.\ J.\ Nucl.\ Phys.\  {\bf 28} (1978) 822.
\bibitem{CCFM}
M.~Ciafaloni,
Nucl.\ Phys.\ {\bf B 296} (1988) 49;\\
%
S.~Catani, F.~Fiorani and G.~Marchesini,
Phys.\ Lett.\ {\bf B 234} (1990) 339;\\
%
S.~Catani, F.~Fiorani and G.~Marchesini,
Nucl.\ Phys.\ {\bf B 336} (1990) 18;\\
G.~Marchesini,
Nucl.\ Phys.\ {\bf B 445} (1995) 49
[hep-ph/9412327].
%
\bibitem{mueller}
A.~H.~Mueller,
Nucl.\ Phys.\ Proc.\ Suppl.\  {\bf 18 C} (1991) 125;\\
%
A.~H.~Mueller,
J.\ Phys.\ {\bf G 17} (1991) 1443.
\bibitem{fpi2}
C.~Adloff {\it et al.}  [H1 Collaboration],
Phys.\ Lett.\ {\bf B 462} (1999) 440
[hep-ex/9907030].
\bibitem{et3}
C.~Adloff {\it et al.}  [H1 Collaboration],
Eur.\ Phys.\ J.\ {\bf C 12} (2000) 595
[hep-ex/9907027].
\bibitem{etzeus}
M.~Derrick {\it et al.}  [ZEUS Collaboration],
Phys.\ Lett.\ {\bf B 356} (1995) 118
[hep-ex/9506012].
\bibitem{emc}
M.~Arneodo {\it et al.}  [European Muon Collaboration],
Phys.\ Lett.\  {\bf B 149} (1984) 415.
%
\bibitem{CTEQ6M}
J.~Pumplin {\it et al.},
JHEP {\bf 0207} (2002) 12
[hep-ph/0201195]. 
\bibitem{sasg}
G.~A.~Schuler and T.~Sj\"ostrand,
Phys.\ Lett.\ {\bf B 376} (1996) 193
[hep-ph/9601282].
\bibitem{lepto} 
G.~Ingelman, A.~Edin and J.~Rathsman,
Comp.\ Phys.\ Comm.\  {\bf 101} (1997) 108
[hep-ph/9605286].
\bibitem{sci} 
G.~Ingelman, A.~Edin and J.~Rathsman,
Phys.\ Lett.\  {\bf B 336} (1996) 371
[hep-ph/9508386].
\bibitem{rapgap}
H.~Jung,
Comp.\ Phys.\ Comm.\  {\bf 86} (1995) 147 \\
(for an update see {\tt http://www-h1.desy.de/\~{ }jung/rapgap/rapgap.html}).
\bibitem{ariadne}
L.~L\"onnblad,
Comp.\ Phys.\ Comm.\  {\bf 71} (1992) 15.
\bibitem{cdm}
B.~Andersson, G.~Gustafson and L.~L\"onnblad,
Nucl.\ Phys.\  {\bf B 339} (1990) 393.
\bibitem{heraws}
N.~H.~Brook {\it et al.} in
{\it Monte Carlo Generators for HERA Physics} (Hamburg, Germany, 1999), 
A.~T.~Doyle, G.~Grindhammer, G.~Ingelman, H.~Jung, Eds.,
 DESY-PROC-1999-02, p.10 [hep-ex/9912053];\\
G.~Grindhammer, D.~Krucker and G.~Nellen, ibid. p.36.
\bibitem{heracles}   
A.~Kwiatkowski, H.~Spiesberger and H.~J.~Mohring,
Comp.\ Phys.\ Comm.\  {\bf 69} (1992) 155.
\bibitem{DJANGO}                                                               K.~Charchula, G.~A.~Schuler and H.~Spiesberger,
Comp.\ Phys.\ Comm.\  {\bf 81} (1994) 381.
\bibitem{cascade}
H.~Jung,
Comp.\ Phys.\ Comm.\  {\bf 143} (2002) 100
[hep-ph/0109102].
\bibitem{cascade1}
H.~Jung and G.~P.~Salam,
Eur.\ Phys.\ J.\ C {\bf 19} (2001) 351
[hep-ph/0012143].
\bibitem{newcascade}
M.~Hansson, H.~Jung,
{\it ``The status of CCFM: Unintegrated gluon densities''}, to 
appear in Proc. of the XI Int. Workshop on Deep Inelastic Scattering
(DIS 2003), St. Petersburg, Russia, April 2003 [hep-ph/0309009].
\bibitem{smallX}   
J.~Andersen {\it et al.}  [Small $x$ Collaboration],
{\it ``Small x phenomenology: Summary and Status 2002''} [hep-ph/0312333].
\bibitem{string}
B.~Andersson, G.~Gustafson, G.~Ingelman and T.~Sj\"ostrand,
Phys.\ Rep.\  {\bf 97} (1983) 31.
\bibitem{jetset74}
T.~Sj\"ostrand,
Comp.\ Phys.\ Comm.\  {\bf 82} (1994) 74;

T.~Sj\"ostrand {\it et al.} 
Comp.\ Phys.\ Comm.\  {\bf 135} (2001) 238
[hep-ph/0010017].
\bibitem{outhwait}
J.~Kwieci\'nski, A.~D.~Martin and J.~J.~Outhwaite,
Eur.\ Phys.\ J.\ {\bf C 9} (1999) 611
[hep-ph/9903439].
\bibitem{pi0frag}
B.~A.~Kniehl, G.~Kramer and B.~P\"otter,
Nucl.\ Phys.\  {\bf B 597} (2001) 337
[hep-ph/0011155].
\bibitem{bfklpar}
A.~D.~Martin, R.~G.~Roberts, W.~J.~Stirling and R.~S.~Thorne,
Eur.\ Phys.\ J.\  {\bf C 4} (1998) 463
[hep-ph/9803445].
\bibitem{con1}
J.~Kwieci\'nski, A.~D.~Martin and P.~J.~Sutton,
Z.\ Phys.\ {\bf C 71} (1996) 585
[hep-ph/9602320].
\bibitem{con2}
J.~Kwieci\'nski, A.~D.~Martin and A.~M.~Sta\'sto,
Phys.\ Rev.\ D {\bf 56} (1997) 3991
[hep-ph/9703445].
\bibitem{aurenche}
P.~Aurenche, R.~Basu, M.~Fontannaz and R.~M.~Godbole,
{\it ``An NLO calculation of the electroproduction of 
large-$E_T$ hadrons''}
[hep-ph/0312359]; \\
M.~Fontannaz, private communication.
\bibitem{MRST99}
A.~D.~Martin, R.~G.~Roberts, W.~J.~Stirling and R.~S.~Thorne,
Eur. Phys. J. {\bf C 23} (2002) 347.
\bibitem{h1det}
I.~Abt {\it et al.}  [H1 Collaboration],
Nucl.\ Instrum.\ Meth.\  {\bf A 386} (1997) 310;

I.~Abt {\it et al.}  [H1 Collaboration],
Nucl.\ Instrum.\ Meth.\  {\bf A 386} (1997) 348.
\bibitem{echpi}
B.~Andrieu {\it et al.}  [H1 Collaboration Calorimeter Group],
Nucl.\ Instrum.\ Meth.\  {\bf A 336} (1993) 499.
\bibitem{enpi0}
B.~Andrieu {\it et al.}  [H1 Collaboration Calorimeter Group],
Nucl.\ Instrum.\ Meth.\  {\bf A 350} (1994) 57.
\bibitem{SPACALTEST}
T.~Nicholls {\it et al.}  [H1 Collaboration SPACAL Group ],
Nucl.\ Instrum.\ Meth.\ {\bf A 374} (1996) 149.
\bibitem{shad}
R.~D.~Appuhn {\it et al.}  [H1 Collaboration SPACAL Group],
Nucl.\ Instrum.\ Meth.\ {\bf A 382} (1996) 395.
\bibitem{glazov}
A.~A.~Glazov,
{\it ``Measurement of the proton structure functions $F_2(x,Q^2)$ and  
$F_L(x,Q^2)$ with the H1 detector at HERA''},
Ph.D. Thesis, Berlin University, 1998, 
DESY-THESIS-1998-005.
\bibitem{twthesis}
T.~Wengler,
{\it ``Measurement of $\pi^\circ$-meson cross sections at low Bjorken-$x$ 
in  deep-inelastic e p collisions at $\sqrt{s}$ = 300 GeV,''}
Ph.D. Thesis, Heidelberg University, 1999, 
DESY-THESIS-1999-011.
\bibitem{lateral}
B.~Andrieu {\it et al.}  [H1 Calorimeter Group Collaboration],
Nucl.\ Instrum.\ Meth.\ {\bf A 344} (1994) 492.
\bibitem{prompt} 
J.~Kwieci\'nski, S.~C.~Lang and A.~D.~Martin,
Phys.\ Rev.\  {\bf D 55} (1997) 1273
[hep-ph/9608355].
\bibitem{phojet}
R.~Engel and J.~Ranft,
Phys.\  Rev.\ {\bf D} 54 (1996) 4244
[hep-ph/9509373].
\bibitem{HFS}
C.~Adloff {\it et al.}  [H1 Collaboration],
Z.\ Phys.\ {\bf C 74} (1997) 221
[hep-ex/9702003].
\bibitem{dijet}   
A.~Aktas {\it et al.}  [H1 Collaboration],
Eur.\ Phys.\ J.\ {\bf C 33} (2004) 477
[hep-ex/0310019].
\bibitem{pjets}
K.~Konishi, A.~Ukawa and G.~Veneziano,
Phys.\ Lett.\ {\bf B} {78} (1978) 243.
%

\end{thebibliography}
\end{document}